\documentclass{article}
\usepackage[T1]{fontenc} 
\usepackage[utf8]{inputenc} 

\usepackage{ismir}
\usepackage{amsmath}
\usepackage{amssymb}
\usepackage{cite}
\usepackage{graphicx}
\usepackage{color}
\usepackage{layouts}
\usepackage{mathtools}
\usepackage{nccmath}

\RequirePackage{algorithm}
\RequirePackage{algorithmic}

\DeclareMathOperator\erf{erf}
\DeclarePairedDelimiter{\abs}{\lvert}{\rvert}

\RequirePackage[hyphens]{url}

\usepackage{lineno}

\title{Symbolic Music Generation with Diffusion Models}

\multauthor
{Gautam Mittal$^{1\star}$\thanks{$^{\star}$ Work completed during an internship at Google Brain.} \hspace{1cm} Jesse Engel$^2$ \hspace{1cm} Curtis Hawthorne$^2$ \hspace{1cm} Ian Simon$^2$} {
 $^1$ University of California, Berkeley \hspace{0.5cm} $^2$ Google Brain\\
{\tt\small gbm@berkeley.edu, \{jesseengel,fjord,iansimon\}@google.com}
}

\def\authorname{G. Mittal, J. Engel, C. Hawthorne, and I. Simon}
\usepackage[bookmarks=false,pdfauthor={\authorname},pdfsubject={\papersubject},hidelinks]{hyperref}

\begin{document}

\maketitle
\begin{abstract}
Score-based generative models and diffusion probabilistic models have been successful at generating high-quality samples in a variety of continuous domains. However, due to their Langevin-inspired sampling mechanisms, their application to discrete symbolic music data has been limited. In this work, we present a technique for training diffusion models on symbolic music data by parameterizing the discrete domain in the continuous latent space of a pre-trained variational autoencoder. Our method is non-autoregressive and learns to generate sequences of latent embeddings through the reverse process and offers parallel generation with a constant number of iterative refinement steps. We show strong unconditional generation and post-hoc conditional infilling results compared to autoregressive language models operating over the same continuous embeddings.
\end{abstract}
\section{Introduction}
\label{submission}




Denoising diffusion probabilistic models (DDPMs)~\cite{pmlr-v37-sohl-dickstein15, ho2020denoising} are a promising new class of generative models that can synthesize comparably high-quality samples by learning to invert a diffusion process from data to Gaussian noise. Unlike many existing deep generative models, DDPMs sample through an iterative refinement process inspired by Langevin dynamics~\cite{welling2011langevin}, which enables post-hoc conditioning of models trained unconditionally~\cite{engel2017latent,song2019generative, song2020improved, du2020implicit} for creative applications. 

Despite these exciting advances, DDPMs have not yet been applied to symbolic music generation because their iterative refinement sampling process is confined to continuous domains such as images~\cite{ho2020denoising} and audio~\cite{chen2020wavegrad, kong2020diffwave}. Similarly, DDPMs cannot take advantage of the recent advances in modeling long-term structure~\cite{razavi2019generating, dhariwal2020jukebox, dalle} that use a two-stage process of modeling discrete tokens extracted by a separate low-level autoencoder. 

In this paper, we demonstrate that it is possible to overcome these limitations by training DDPMs on the continuous latents of a low-level variational autoencoder (VAE) to generate long-form discrete symbolic music. Our key findings include:

\begin{itemize}
\itemsep0em
\item High-quality unconditional sampling of discrete melodic sequences (1024 tokens) with DDPMs through iterative refinement of lower-level VAE latents.
\item DDPMs outperforming strong autoregressive baselines (TransformerMDN) in hierarchical modeling of continuous latents, partly due to a lack of teacher forcing and exposure bias during training.
\item Post-hoc conditional infilling of melodic sequences for creative applications. 


\end{itemize}

\section{Background}

\begin{figure*}[ht]
\vskip 0in
\begin{center}
\centerline{\includegraphics[width=0.7\textwidth]{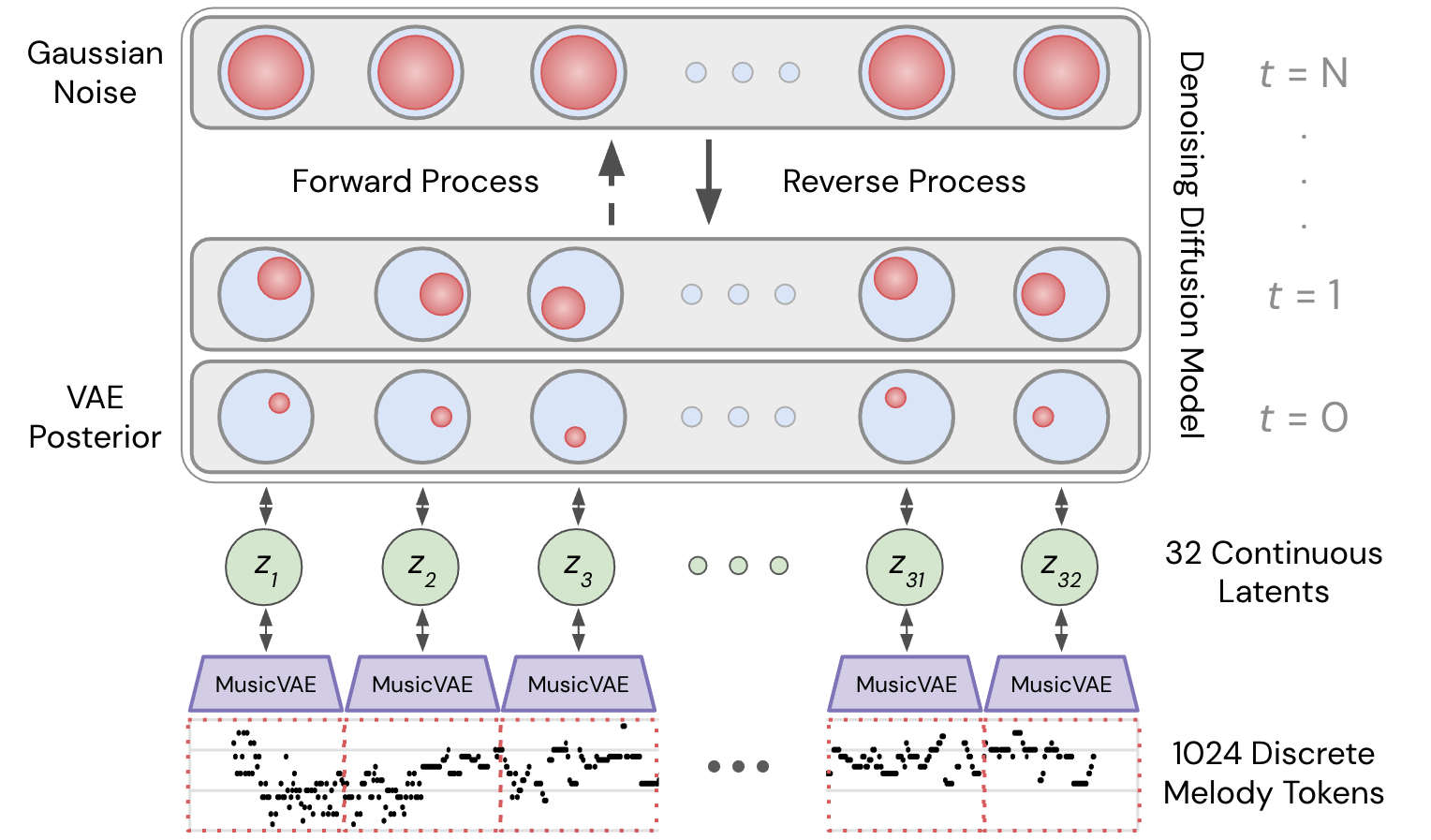}}
\caption{A diagram of our proposed framework. We use the pre-trained 2-bar melody MusicVAE~\cite{roberts2019hierarchical} to embed discrete musical phrases (64 bars, 1024 tokens) into a sequence of continuous latent codes (32 latents, 512 dimensions each). These embeddings are used to train a diffusion model that iteratively adds noise such that after $N$ diffusion steps the input embeddings are distributed $\mathcal{N}(0, I)$. To sample from this model, we initialize Gaussian noise and use the reverse process to iteratively refine the noise samples into a sequence of embeddings from the data distribution. These generated embeddings are fed through the MusicVAE decoder to produce the final MIDI sequence.}
\label{mainarch}
\end{center}
\vskip -0.2in
\end{figure*}

\subsection{Denoising Diffusion Probabilistic Models}

DDPMs~\cite{pmlr-v37-sohl-dickstein15, ho2020denoising} are a class of generative models that define latents $x_1, ..., x_N$ of the same dimensionality as the data $ x_0 \sim q(x_0) $. Diffusion models are comprised of a \textbf{forward process} and a \textbf{reverse process}. The \textbf{forward process} starts from the data $ x_0 $ and iteratively adds Gaussian noise according to a fixed noise schedule for $N$ diffusion steps:
\begin{gather}
q(x_t|x_{t-1}) = \mathcal{N}(x_t; \sqrt{1 - \beta_{t}}x_{t-1}, \beta_{t}I) \\
q(x_{1:N} |x_0) = \prod_{t=1}^{N} q(x_t|x_{t-1})
\end{gather}  
where $\beta_{1}, \beta_{2}, ..., \beta_{N}$ is a noise schedule that converts the data distribution $x_0$ into latent $x_N$. 
The choice of noise schedule has been shown to have important effects on sampling efficiency and quality~\cite{chen2020wavegrad, ho2020denoising}.

The \textbf{reverse process} is defined by a Markov chain parameterized by $\theta$ that iteratively refines latent point $x_N \sim \mathcal{N}(0, I)$ into data point $x_0$.
The learned transition probabilities are defined as,
\begin{gather}
p_{\theta}(x_{t-1}|x_t) = \mathcal{N}(x_{t-1}; \mu_{\theta}(x_t, t), \sigma_{\theta}(x_t, t)) \\
p_{\theta}(x_{0:N}) = p(x_N) \prod_{t=1}^{N} p_{\theta}(x_{t-1}|x_t)
\end{gather}
where the objective is to gradually denoise samples at each reverse diffusion step $t$.
In practice, $\sigma_\theta $ is set to an untrained time-dependent constant based on the noise schedule, and~\cite{ho2020denoising} found $\sigma_\theta(x_t, t) = \sigma_t = \frac{1 - \bar{\alpha}_{t-1}}{1 - \bar{\alpha}_t} \beta_t$ to have reasonable practical results, where $\alpha_t = 1 - \beta_t$, and $\bar{\alpha}_t = \prod_{i=1}^{t} \alpha_i$.

The training objective is to maximize the log likelihood of $ p_\theta(x_0) = \int p_\theta(x_0,...,x_N)dx_{1:N}$, but the intractability of this marginalization leads to the following evidence lower bound (ELBO):
\begin{gather}
\begin{split}
    \mathbb{E}\left[\log p_\theta (x_0)\right] \geq 
    \mathbb{E}_{q}\left[\log \frac{p_\theta(x_{0:N})}{q(x_{1:N}|x_0)}\right] \\
    = \mathbb{E}_{q}\left[
        \log p(x_N) + 
        \sum_{t \geq 1} \log \frac{p_\theta(x_{t-1}|x_t)}{q(x_t|x_{t-1})}
    \right]
\end{split}
\end{gather}
Additionally,~\cite{ho2020denoising} observed that the forward process can be computed for any step $t$ such that $q(x_t | x_0) = \mathcal{N}(x_{t}; \sqrt{\bar{\alpha}_t}x_0, (1-\bar{\alpha}_t)I)$, which can be viewed as a stochastic encoder. To simplify the above variational bound,~\cite{ho2020denoising} propose training on pairs of $(x_t, x_0)$ to learn to parameterize this process with a simple squared L2 loss. The following objective is simpler to train, resembles denoising score matching~\cite{vincent2011connection, song2019generative} and was found to yield higher-quality samples:
\begin{gather}
    L(\theta) = \mathbb{E}_{x_0, \epsilon, t}\left[\left\Vert 
    \epsilon - \epsilon_\theta(\sqrt{\bar{\alpha}_t}x_0 + \sqrt{1-\bar{\alpha}_t}\epsilon, t)
    \right\Vert^2 \right]
\end{gather}
where $ t $ is sampled uniformly between $1$ and $N$, $\epsilon \sim \mathcal{N}(0, I)$, and $\epsilon_\theta$ is the learned diffusion model.~\cite{chen2020wavegrad} found that instead of conditioning on a discrete diffusion step $t$, it was beneficial to sample a continuous noise level $\sqrt{\bar{\alpha}} \sim \mathbb{U}(\sqrt{\bar{\alpha}_{t-1}}, \sqrt{\bar{\alpha}_t})$ where $t \sim \mathbb{U}(\{1,...,N\})$ and $\bar{\alpha}_0 = 1$.

We refer readers to Algorithms~\ref{alg:ddpm_train} and~\ref{alg:ddpm_sample} based on~\cite{chen2020wavegrad} in our supplementary material\footnote{Supplementary material including additional implementation details, audio, and tables is available at \url{https://goo.gl/magenta/symbolic-music-diffusion-examples}} for the full training and sampling procedure.



\subsection{Variational Autoencoders}

Variational autoencoders~\cite{kingma2014auto} are generative models that define $ p(y, z) = p(y|z)p(z) $ where $z$ is a learned latent code for data point $y$. Additionally, the latent code is constrained such that $z$ is distributed according to prior $p(z)$ where the prior is usually an isotropic Gaussian. The VAE is comprised of an encoder $q_\gamma(z|y)$ which models the approximate posterior $p(z|y)$ and a decoder $p_\theta(y|z)$ which models the conditional distribution of data $y$ given latent code $z$. 

The training objective is to maximize the log likelihood of $p_\theta(x) = \int p_\theta(y|z)p(z) dz$, but this marginalization is intractable, and we use the following variational bound maximized with $q_\gamma(z|y)$ as the approximate posterior:
\begin{equation}
    \mathbb{E} \left[ \log p_\theta(y|z) \right] 
    - \mathrm{KL}(q_\gamma(z|y) || p(z)) \leq \log p(y)
\end{equation}
The flexible implementation of variational autoencoders allows them to learn representations over a wide variety of domains. Of particular interest to us are sequential autoencoders~\cite{bowman2016generating, roberts2019hierarchical} which use long short-term memory cells~\cite{hochreiter1997long} to model temporal context in sequential data distributions.

In practice, there is a trade-off between the quality of reconstructions and the distance between the approximate posterior $q_\gamma(z|y)$ and the Gaussian prior $p(z)$. This makes sampling more difficult for VAEs with better reconstructions due to latent ``holes" in the approximate posterior and is one of the primary shortcomings of these models.
\section{Model}
A diagram and description of our multi-stage diffusion model is shown in Figure~\ref{mainarch}. We refer readers to the supplement for full implementation details.

\subsection{Architecture}
Our model learns to generate discrete sequences of notes (known as MIDI) by first training a VAE with parameters $\gamma$ on the sequences and then training a diffusion model to capture the temporal relationships among the $k$ VAE latents.
Sequence VAEs such as MusicVAE are difficult to train on long sequences~\cite{roberts2019hierarchical}, which we overcome by pairing the short 2-bar MusicVAE model with a diffusion model capable of modeling dependencies between $k=32$ latents, thus modeling 64 bars in total.

\textbf{MusicVAE embeddings: } Each musical phrase is a sequence of one-hot vectors with 16 quantized steps per measure and the vocabulary contains 90 possible tokens (1 note on + 1 note off + 88 pitches). We then parameterize each 2-bar phrase using the pre-trained 2-bar melody MusicVAE~\cite{roberts2019hierarchical} and generate a sequence of continuous latent embeddings $z_1,...,z_k$ to parameterize an entire sequence. The MusicVAE model employs bidirectional recurrent neural networks as an encoder and autoregressive decoding as shown in Figure~\ref{musicvae-architecture} in the supplement. As we use the pretrained model from the original work, full model details can be found in~\cite{roberts2019hierarchical}. After encoding each 2-measure phrase into a latent $z$ embedding, we perform linear feature scaling such that the domain of each embedding is $[-1, 1]$. This ensures consistently scaled inputs starting from the isotropic Gaussian latent $x_N$ for the diffusion model.

\textbf{Transformer diffusion model:} Our network $\epsilon_\theta(x_t, \sqrt{\bar{\alpha}}): \mathbb{R}^{k \times 42} \times \mathbb{R} \rightarrow \mathbb{R}^{k \times 42}$ is a transformer~\cite{vaswani2017attention} where $k = 32$ is the length of each sequence of 42-dimensional preprocessed latent embeddings. The unperturbed data distribution used to train the diffusion model is $x_0 = [z_1, ..., z_k]$. The network contains an initial fully-connected layer that projects the embeddings into a 128-dimensional space, followed by $L=6$ encoder layers each with $H = 8$ self-attention heads and a residual fully-connected layer. All self-attention and fully-connected layers use layer normalization \cite{ba2016layer}. The output of the encoder is fed to $K = 2$ noise-conditioned residual fully-connected layers which generate the reverse process output. Each fully-connected layer contains 2048 neurons. We use a 128-dimensional sinusoidal positional encoding similar to~\cite{vaswani2017attention} where $j$ is the position index of a latent input embedding:
\begin{equation}
\begin{split}
    \omega = \left[ 10^{\frac{-4 \times 0}{63}}j,...,10^{\frac{-4 \times 63}{63}}j \right] \hspace{1mm}
    e_j= \left[ \sin(\omega),  \cos(\omega) \right]
\end{split}
\end{equation}
This positional encoding $e_1, e_2, ..., e_k$ is added to inputs $x_t$ before being fed through the transformer encoder layers allowing the model to capture the temporal context of the continuous inputs.

\textbf{Noise schedule and conditioning:} As described in both the original diffusion model framework~\cite{ho2020denoising} and in~\cite{chen2020wavegrad}, we use an additional sinusoidal encoding to condition the diffusion model on a continuous noise level during training and sampling. This noise encoding is identical to the positional encoding described above but with the frequency of each sinusoid scaled by 5000 to account for the updated domain. We use feature-wise linear modulation~\cite{perez2018film} to generate $\gamma$ (scale) and $\xi$ (shift) parameters given a noise encoding and apply the transformation $\gamma \phi + \xi$ to the output $\phi$ of each layer normalization block in each residual layer, allowing for effective conditioning of the diffusion model. Our model uses a linear noise schedule with $N = 1000$ steps and $\beta_1 = 10^{-6}$ and $\beta_N = 0.01$.

\subsection{Unconditional Generation}
In the unconditional generation task, the goal is to produce samples that exhibit long-term structure. Because of our multi-stage approach, this works even in the scenario where the KL divergence between the marginal posterior $q_\gamma(z)$ and the Gaussian prior is quite large because the diffusion model accurately captures the structure of the latent space therefore improving the sample quality. Additionally, we extend the underlying VAE to samples longer than what it was trained to model by using the diffusion model to predict sequences of latent embeddings and attempt to generate unconditional samples with coherent patterns across a large number of measures.

\subsection{Infilling}
One of the benefits of using a sampling process that iteratively refines noise into data samples is that the trajectory of the reverse process can be steered and arbitrarily conditioned without the need for retraining the diffusion model. In creative domains, this post-hoc conditioning is especially useful for artists without the computational resources to modify or re-train deep models for new tasks. We demonstrate the power of diffusion modeling applied to music with conditional infilling of latent embeddings using an unconditionally trained diffusion model. 

The infilling procedure extends the sampling procedure described in Algorithm~\ref{alg:ddpm_sample} by incorporating information from a partially occluded sample $s$. At each step of sampling, we diffuse the fixed regions of $s$ with the forward process $q(s_t|s) = \mathcal{N}(s_{t}; \sqrt{\bar{\alpha}_t}s, (1-\bar{\alpha}_t)I)$ and use a mask $m$ to add the diffused fixed regions to the updated sample $x_{t-1}$. The final output $x_0$ will be a version of $s$ with the occluded regions inpainted by the reverse process.

We refer readers to Algorithm~\ref{alg:infilling} for the modified sampling procedure that allows for post-hoc conditional infilling. 

\begin{algorithm}[H]
   \caption{Infilling}
   \label{alg:infilling}
\begin{algorithmic}
   \STATE {\bfseries Input:} mask $m$, sample $s$, $N$ steps, $\beta_1, ..., \beta_N$
   \STATE $x_N \sim \mathcal{N}(0, I) $
   \FOR{$t=N, . . .,1$}
   \STATE $\epsilon_1, \epsilon_2 \sim \mathcal{N} (0, I)$ if $t > 1$, else $\epsilon_1 = \epsilon_2 = 0$
   \STATE $ y = \sqrt{\bar{\alpha}_t}s + \sqrt{1-\bar{\alpha}_t}\epsilon_1$ if $t > 1$, else $s$
   \STATE $x_{t-1} = \frac{1}{\sqrt{\alpha_t}}
        \left(x_t - \frac{1 - \alpha_t}{\sqrt{1 - \bar{\alpha}_t}}
                \epsilon_\theta(x_t, \sqrt{\bar{\alpha}_t})
        \right) + \sigma_t \epsilon_2$
    \STATE $x_{t-1} = x_{t-1} \odot (1 - m) + y \odot m$
   \ENDFOR
   \STATE \textbf{return} $x_0$
\end{algorithmic}
\end{algorithm}
\section{Methods}

\begin{figure*}[ht]
\vskip 0in
\begin{center}
\centerline{\includegraphics[width=\textwidth]{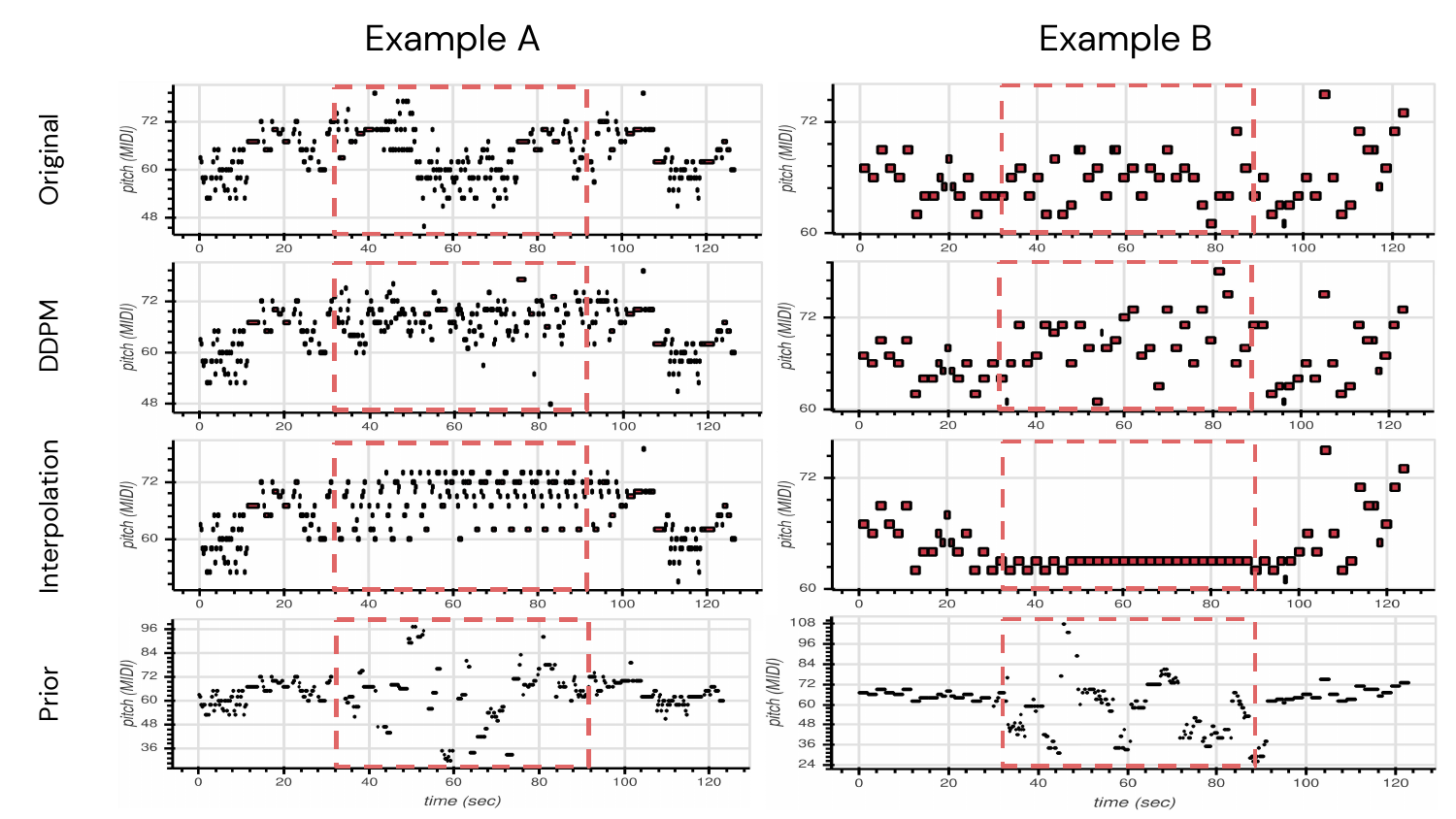}}
\caption{Example piano rolls of infilling experiments. For each example (A, B) the first and last 256 melody tokens are held constant, and the interior 512 tokens are filled in by the model (dashed red box). Even qualitatively, it is visually apparent that the diffusion model (second row) produces notes with a consistency and variance similar to the original data (first row), while the latent interpolation (third row) is too repetitive, and sampling independently from the prior (last row) produces outputs with too much variety and lack of local coherence.}
\label{fig:infilling-example}
\end{center}
\vskip -0.2in
\end{figure*}

\subsection{Data}
We use the Lakh MIDI Dataset (LMD)~\cite{raffel2016learning} for all experiments. The dataset contains over 170,000 MIDI files with 99\% of those files used for training and the remaining used for validation. We extracted 988,893 64-bar monophonic sequences for training and 11,295 for validation from the provided MIDI files. Each sequence was encoded into 32 continuous latent embeddings using MusicVAE. We set the softmax temperature for MusicVAE to 0.001 for decoding generated embeddings in all experiments. A diagram of the MusicVAE architecture used is shown in the supplementary material.

\subsection{Autoregressive Baseline}
We compare our model to an autoregressive transformer with a mixture density output layer~\cite{bishop1994mixture} and train on the same dataset as the diffusion model. To ensure a fair comparison, we use the same architecture as our diffusion model with $L = 6$, $ H = 8 $, and $K = 2$ with 2048 neurons for each fully-connected layer. The mixture density layer outputs a mixture of 100 Gaussians to ensure sufficient mode coverage. In total, our baseline model has 38M trainable parameters. While the model is the same as the diffusion model (25.58M trainable parameters) up until the output layer, the autoregressive model has more parameters due to the much larger output layer. We refer to this model as TransformerMDN.

\subsection{Training}
All models were trained using Adam~\cite{kingma2014adam} with default parameters. We trained our diffusion model for 500K steps on a single NVIDIA Tesla V100 GPU for 6.5 hours using a learning rate of $10^{-3}$ annealed with a decay rate of 0.98 every 4000 steps and batch size 64. Unlike the diffusion model, which is non-autoregressive, we train TransformerMDN with teacher forcing. We use a batch size of 128, learning rate $3\times10^{-4}$, and train for 250K steps on a single NVIDIA Tesla V100 GPU for 6.5 hours.


We used the open-source implementation of MusicVAE written in TensorFlow~\cite{abadi2016tensorflow} and the publicly available 2-bar melody checkpoints trained on LMD. We trained our diffusion and baseline models\footnote{Our implementation is available at \url{https://github.com/magenta/symbolic-music-diffusion}} with JAX~\cite{jax2018github} and Flax~\cite{flax2020github}.

\subsection{Framewise Self-similarity Metric}
To evaluate the statistical similarity between our model's qualitative output and the original training sequences, we present a metric that captures local self-similarity patterns across generated melodic sequences. Inspired by the statistical similarity evaluation described in~\cite{choi2020encoding}, we evaluate our models with a modified framewise Overlapping Area ($\mathrm{OA}$) metric. 

We use a sliding 4-measure window with 2-measure hop size to capture local pitch and duration statistics across the piece. Within each 4-measure frame, we compute the mean and variance of both pitch, which captures melodic similarity, and duration, which captures rhythmic similarity. These statistics specify a Gaussian PDF for pitch and duration for each frame $(p_P(k), p_D(k))$. We compute the Overlapping Area ($\mathrm{OA}$) \cite{choi2020encoding} of adjacent frames ($k$, $k+1$) where each frame's statistics are modeled as $\mathcal{N}(\mu_1, \sigma^2_1)$ and $\mathcal{N}(\mu_2, \sigma^2_2)$, respectively:
\begin{equation}
  \mathrm{OA}(k, k+1) = 1 - \erf \left( \frac{c - \mu_1}{\sqrt{2}\sigma^2_1} \right) + \erf \left( \frac{c - \mu_2}{\sqrt{2}\sigma^2_2} \right)
\end{equation}
for both pitch ($\mathrm{OA}_P$) and duration ($\mathrm{OA}_D$), where $\erf$ is the Gauss error function and $c$ is the point of intersection between Gaussian PDFs with $\mu_1 < \mu_2$. 
For a set of MIDI samples, we infer the  \textit{Consistency} and \textit{Variance} from the mean ($\mu_{\mathrm{OA}}$) and variance ($\sigma^2_{\mathrm{OA}}$) respectively of $\mathrm{OA}$ aggregated over all adjacent frames.
We then use these aggregate values to compute the normalized relative similarity of pitch and duration consistency and variance to the training set (GT):
\begin{equation}
\begin{split}
    Consistency &= \max(0, 1 - \frac{\abs{\mu_{\mathrm{OA}} - \mu_{GT}}}{\mu_{GT}}) \\
    Variance &= \max(0, 1 - \frac{\abs{\sigma^2_{\mathrm{OA}} - \sigma^2_{GT}}}{\sigma^2_{GT}})
\end{split}
\label{eq:convar}
\end{equation}
We clip \textit{Consistency} and \textit{Variance} such that samples with $\mu_{OA}$ or $\sigma^2_{OA}$ with greater than 100\% percent error from the ground truth are considered to have zero relative similarity.

\subsection{Latent Space Evaluation}
\label{sec:latent}
We evaluate the similarity of latent embeddings generated by each of our models using the Fr\'echet distance (FD)~\cite{heusel2018gans} and Maximum Mean Discrepancy (MMD)~\cite{JMLR:v13:gretton12a} with a polynomial kernel, which are popular evaluation metrics in the generative modeling literature. These metrics measure the distance between the models' continuous output distributions and the original data distribution in latent space. It is important to note that this metric does not measure long-term temporal consistency or quality of produced sequences and only measures the quality of the intermediate continuous representation before the final sequence is generated using the MusicVAE decoder. 

\section{Results}

\begin{table}[tb]
\begin{center}
\resizebox{\columnwidth}{!}{%
\begin{tabular}{|l|c|c|c|c|c|c|c|c|r|}
\hline
Setting      & \multicolumn{4}{c}{Unconditional} \vline & \multicolumn{4}{c}{Infilling} \vline \\ 
\hline
Quantity      & \multicolumn{2}{c}{Pitch} \vline & \multicolumn{2}{c}{Duration}  \vline & \multicolumn{2}{c}{Pitch} \vline & \multicolumn{2}{c}{Duration} \vline \\ 
\hline
 Metric & C & Var & C & Var & C & Var & C & Var    \\
\hline
Train Data            & 1.00 & 1.00 & 1.00 & 1.00 & 1.00 & 1.00 & 1.00 & 1.00\\
Test Data              & 1.00 & 0.96 & 1.00 & 0.91 & 1.00 & 0.96 & 1.00 & 0.91\\
\hline
Diffusion  & \textbf{0.99} & \textbf{0.90} & \textbf{0.96} & \textbf{0.92} & \textbf{0.97} & \textbf{0.87} & \textbf{0.97} & \textbf{0.80} \\
Autoregression    & 0.93 & 0.68 & 0.93 & 0.76 & - & - & - & -\\
Interpolation     & 0.85 & 0.23 & 0.91 & 0.34 & 0.94 & 0.78 & 0.96 & \textbf{0.80} \\
$\mathcal{N}(0, I)$ Prior   & 0.84 & 0.19 & 0.90 & 0.67  & 0.89 & 0.19 & 0.94 & 0.54\\

\hline
\end{tabular}
}
\end{center}
\caption{Framewise self-similarity of consistency (C) and variance (Var), as defined in Equation~\ref{eq:convar}, for note pitch and duration. For both unconditional sampling and infilling tasks, the diffusion model produces samples most similar to the real data. For diffusion samples, we use $N = 1000$ sampling steps with $\beta_1 = 10^{-6}$ and $\beta_N = 0.01$. For the TransformerMDN baseline we sample with a temperature of 1.0, meaning we sampled directly from the logits of mixture density layer. Absolute values of overlap area can be found in Table~\ref{abs-value-table} in our supplementary material.}
\label{rel-table}
\end{table}

\subsection{Unconditional Generation} 

\begin{figure}[ht]
\vskip 0in
\begin{center}
\centerline{\includegraphics[width=0.89\columnwidth]{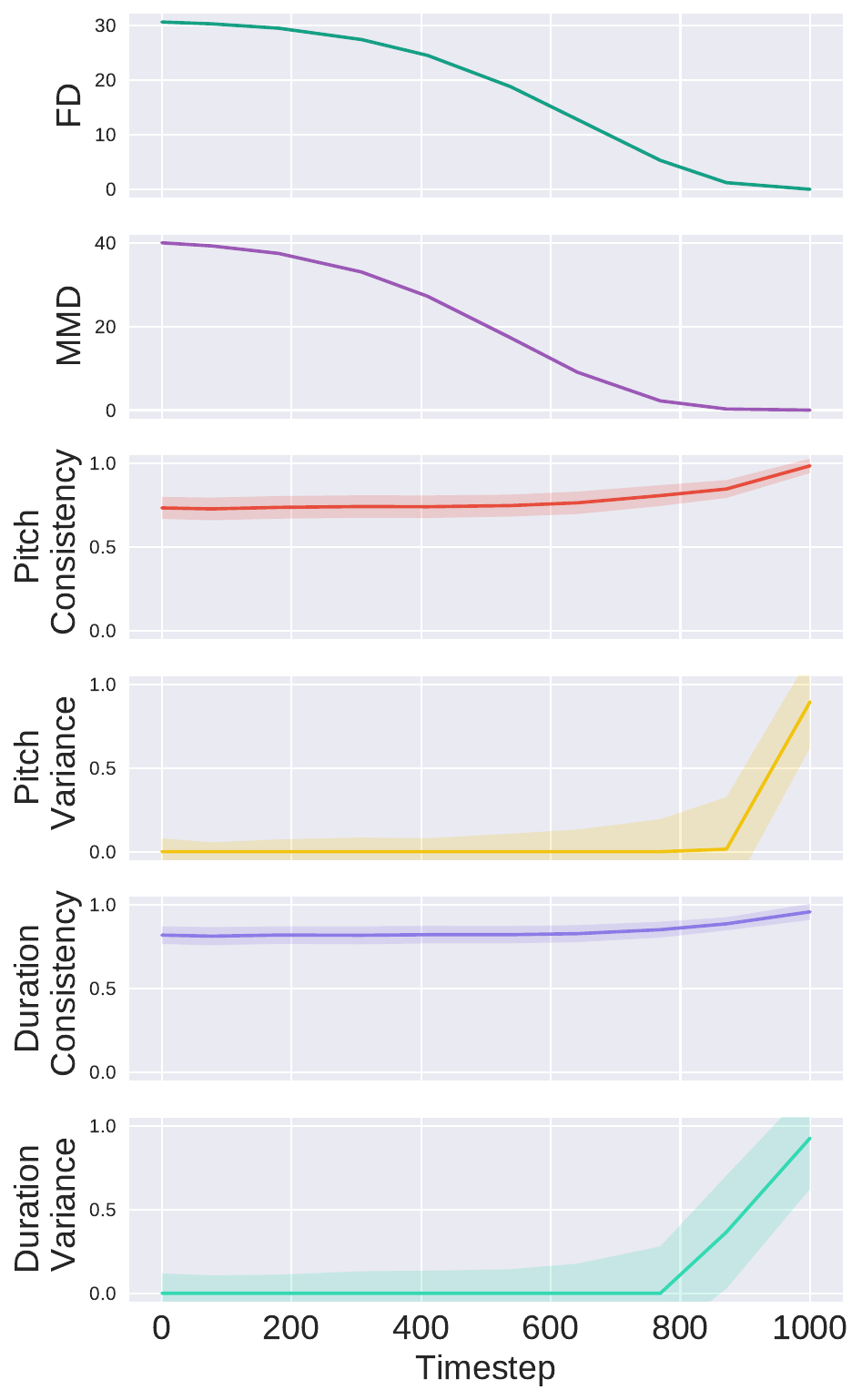}}
\caption{Sample quality improvement during iterative refinement. Latent space and framewise metrics evaluated at different stages of unconditional sampling. The metrics improve as the iterative refinement process progresses. We plot the means and standard deviations for 1000 samples.}
\label{fig:inference-curve}
\end{center}
\vskip -0.3in
\end{figure}
To evaluate unconditional sample quality, we compare batches of 1000 samples (32 latents each) generated by our proposed diffusion model with random draws from the training and test sets. We compare against a set of baseline generators including TransformerMDN, the independent $\mathcal{N}(0, I)$ MusicVAE prior, and spherical interpolation~\cite{white2016sampling} between two MusicVAE embeddings at the start and end of an example from the test set. 

As seen in Table~\ref{rel-table}, the diffusion model quantitatively produces samples most similar to the training data according to the relative framewise overlapping area metrics for note pitch and duration. The diffusion model outperforms TransformerMDN, which is challenged by modelling the relatively high-dimensional continuous latents autoregressively, even with a mixture of Gaussians output. The autoregressive models are also trained with teacher forcing that results in exposure bias, leading to divergence from the data distribution during sampling. This is reflected in the lower pitch and duration consistency and higher variance in the absolute overlapping area numbers seen in Supplemental Table~\ref{abs-value-table}. Additionally, the diffusion model is able to capture the joint dependencies of the sequences better because it learns to model all latents simultaneously as opposed to autoregressively. Note that the Gaussian prior also suffers from low consistency and high variance, due to lack of temporal dependencies, while the interpolated samples conversely suffer from low variance and too much consistency due to high repetition.

Table~\ref{latent-table} in our supplement presents latent space evaluations of our generated samples. The TransformerMDN outperformed all other models, likely due to the Gaussian mixture prior on its output layer whereas the diffusion model must learn the output distribution from scratch. Furthermore, the latent space metrics are limited by assumptions about the latent manifold distribution and are unable to fully capture the detail of the entire space, further highlighting the necessity of our quantitative framewise self-similarity metrics and qualitative evaluations.

Figure~\ref{fig:inference-curve} helps us to further understand the iterative refinement process by showing the improvement in sample quality as a function of iterative refinement time for both latent space and framewise self-similarity metrics. Interestingly, latent metrics improve steadily, while consistency similarity starts fairly high, and variance similarity only emerges at the end of refinement. We refer the reader to the supplementary material for extended visual and audio samples of the generated sequences from each model\footnote{\url{https://goo.gl/magenta/symbolic-music-diffusion-examples}}.


\subsection{Infilling}
To probe the diffusion model's ability to perform post-hoc conditional generation, we remove the middle 32 measures (16 embeddings) and generate new embeddings following Algorithm~\ref{alg:infilling} by conditioning on the first and last 8 embeddings. As in the unconditional setting, we compare to interpolation and independent samples from the prior.

Figure~\ref{fig:infilling-example} visualizes the task by plotting the resulting note sequences for two different examples. Even qualitatively, it is visually apparent that the diffusion model produces notes with a consistency and variance similar to the original data. Latent interpolation is very consistent but unrealistically repetitive, and sampling independently from the prior produces a sequence with extremely large variance that is inconsistent in both pitch and duration.

The quantitative evaluations in Table~\ref{rel-table} back up these observations. Similar to the unconditional generation task, the diffusion model outperforms the baselines in both consistency and variance similarity. We do not include the autoregressive baseline because it is unable to condition on the final 8 embeddings. 


\section{Related Work}

\textbf{Multi-stage learning:} Several models have achieved long-term structure by first modeling some intermediate representation and then using that to guide the final generative process. Wave2Midi2Wave~\cite{hawthorne2018enabling} uses a transformer to generate MIDI-like symbolic data and then a WaveNet~\cite{oord2016wavenet} to synthesize that symbolic data into audio. Jukebox~\cite{dhariwal2020jukebox} and DALL-E~\cite{dalle} use similar approaches for text-conditioned generative music audio and image models. There has also been work investigating the extension of a single-measure VAE to multiple measures by using an autoregressive LSTM with a mixture density output layer~\cite{Jhamtani2019ModelingSI}, similar to TransformerMDN.


\textbf{Iterative refinement:}~\cite{huang2017coconet} use an orderless NADE with blocked Gibbs sampling to iteratively rewrite musical harmonies based on surrounding context.~\cite{lattner2016imposing} use a gradient-based sampler combined with a restricted Boltzmann machine to generate polyphonic piano music. Similarly,~\cite{zhang2020bachgrad} investigated the use of a score-based generative model~\cite{song2019generative} to generate Bach chorales with annealed Langevin dynamics. A key distinction between our method and prior work is the use of a VAE to parameterize the discrete space of musical notes for improved generation with a DDPM while previous methods have performed iterative refinement in the discrete space directly.

\textbf{Conditional sampling from unconditional models:} We build on top of the breadth of material that investigate steering generation in the latent space of an unconditionally trained generative model~\cite{nguyen2017plug, jaques2018learning, jahanian2019steerability, dathathri2019plug}. Most similar to our work is~\cite{nguyen2017plug, engel2017latent, midime} which train an additional model on top of a pre-trained variational autoencoder to steer generation in the latent space of that autoencoder. Our approach builds on top of this by not only improving sampling and generation of the underlying autoencoder but also extending generation to sequences much longer than those used to train the VAE. We also use a diffusion model to refine generation and provide conditional infilling while the works mentioned primarily used conditional GANs and VAEs to extend the underlying autoencoder for a single latent embedding as opposed to a sequence of latent embeddings.


\section{Conclusion}
We have proposed and demonstrated a multi-stage generative model comprised of a low-level variational autoencoder with continuous latents modeled by a higher-level diffusion model. This approach enables using diffusion models on discrete data, and as priors for modeling long-term structure in multi-stage systems.
We demonstrate that this model is useful for symbolic music generation, both in unconditional generation and conditional infilling.
Future work will include extending this approach to other discrete data such as text, and exploring a greater array of approaches for post-hoc conditioning in creative applications.
\section{Acknowledgements}
We thank Anna Huang, Carrie Cai, Sander Dieleman, Doug Eck and the rest of the Magenta team for helpful discussions, feedback, and encouragement. 

\bibliography{references}

\newpage
\appendix

\section{DDPM Training and Sampling}
To train and sample from our diffusion model, we use the algorithms as described in~\cite{chen2020wavegrad}.

\begin{algorithm}[H]
   \caption{Training}
   \label{alg:ddpm_train}
\begin{algorithmic}
   \STATE {\bfseries Input:} $q(x_0)$, $N$ steps, noise schedule $\beta_1, ..., \beta_N$
   \REPEAT
   \STATE $x_0 \sim q(x_0)$
   \STATE $t \sim \mathbb{U}(\{1,...,N\})$
   \STATE $\sqrt{\bar{\alpha}} \sim \mathbb{U}(\sqrt{\bar{\alpha}_{t-1}}, \sqrt{\bar{\alpha}_t}) $
   \STATE $\epsilon \sim \mathcal{N}(0, I)$
   \STATE Take gradient descent step on
   \STATE $\nabla_\theta \left\Vert 
    \epsilon - \epsilon_\theta(\sqrt{\bar{\alpha}_t}x_0 + \sqrt{1-\bar{\alpha}_t}\epsilon, \sqrt{\bar{\alpha}})
    \right\Vert^2$
   \UNTIL{converged}
\end{algorithmic}
\end{algorithm}

\begin{algorithm}[H]
   \caption{Sampling}
   \label{alg:ddpm_sample}
\begin{algorithmic}
   \STATE {\bfseries Input:} $N$ steps, noise schedule $\beta_1, ..., \beta_N$
   \STATE $x_N \sim \mathcal{N}(0, I) $
   \FOR{$t=N, . . .,1$}
   \STATE $\epsilon \sim \mathcal{N} (0, I)$ if $t > 1$, else $\epsilon = 0$
   \STATE $x_{t-1} = \frac{1}{\sqrt{\alpha_t}}
        \left(x_t - \frac{1 - \alpha_t}{\sqrt{1 - \bar{\alpha}_t}}
                \epsilon_\theta(x_t, \sqrt{\bar{\alpha}_t})
        \right) + \sigma_t \epsilon$
   \ENDFOR
   \STATE \textbf{return} $x_0$
\end{algorithmic}
\end{algorithm}

\section{MusicVAE Architecture}

\begin{figure}[ht]
\begin{center}
\centerline{\includegraphics[width=.6\columnwidth]{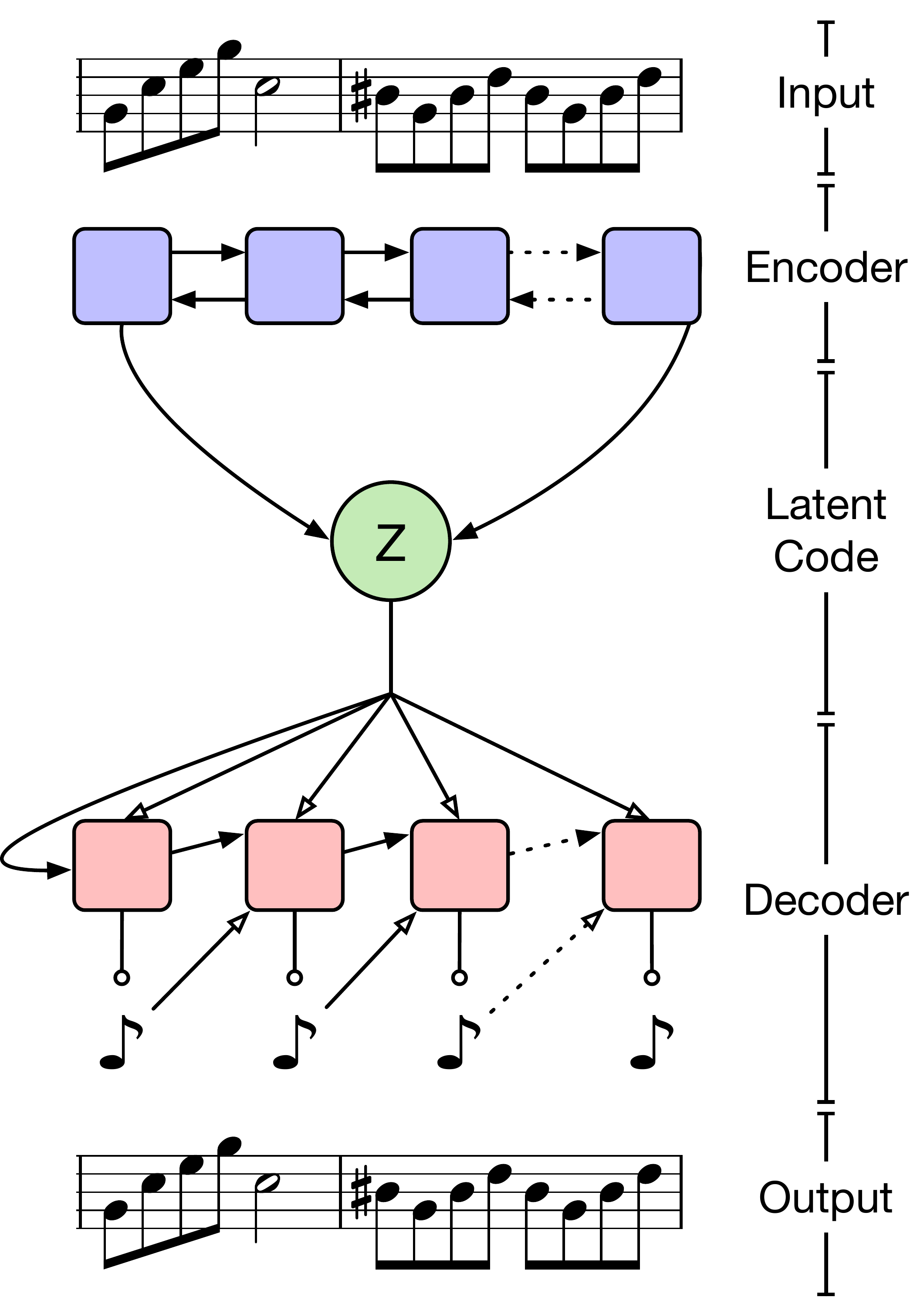}}
\caption{2-bar melody MusicVAE architecture. The encoder is a bi-direction LSTM and the decoder is an autoregressive LSTM.}
\label{musicvae-architecture}
\end{center}
\vskip -0.2in
\end{figure}

\section{Trimming Latents}

\begin{figure}[H]
\vskip 0in
\begin{center}
\centerline{\includegraphics[width=\columnwidth]{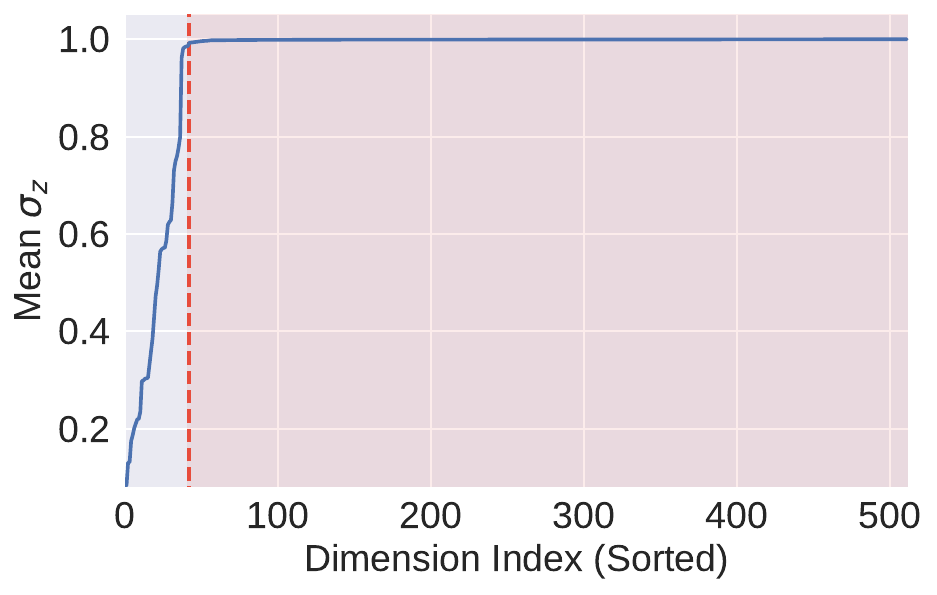}}
\caption{The standard deviation per dimension of the MusicVAE posterior $q(z|y)$ averaged across the entire training set. The region highlighted in red contains the latent dimensions that are unused.}
\label{vae-latent-analysis}
\end{center}
\vskip -0.2in
\end{figure}

During training VAEs typically learn to only utilize a fraction of their latent dimensions.
As shown in Figure~\ref{vae-latent-analysis}, by examining the standard deviation per dimension of the posterior $q(z|y)$ averaged across the entire training set, we are able to identify underutilized dimensions where the average embedding standard deviation is close to the prior of 1.
The VAE loss encourages the marginal posterior to match to the prior~\cite{hoffman2016elbo, alemi2018fixing}, but to encode information, dimensions must have smaller variance per an example.

In all experiments, we remove all dimensions except for the 42 dimensions with standard deviations below 1.0, before training the diffusion model on the input data. We find this latent trimming to be essential for training as it helps to avoid modeling unnecessary high-dimensional noise and is very similar to the distance penalty described in~\cite{engel2017latent}. We also tried reducing the dimensionality of embeddings with principal component analysis (PCA) but found that the lower dimensional representation captured too many of the noisy dimensions and not those with high utilization.

\section{Tables}
In Tables \ref{abs-value-table} and \ref{latent-table}, we present the unnormalized framewise self-similarity results as well as the latent space evaluation of each model.

\begin{table}[h]
  \begin{center}
  \resizebox{\columnwidth}{!}{%
  \begin{tabular}{|l|c|c|c|c|c|c|c|c|}
  \hline
  Setting      & \multicolumn{4}{c}{Unconditional} \vline & \multicolumn{4}{c}{Infilling} \vline \\ 
  \hline
  Quantity      & \multicolumn{2}{c}{Pitch} \vline & \multicolumn{2}{c}{Duration} \vline & \multicolumn{2}{c}{Pitch} \vline & \multicolumn{2}{c}{Duration} \vline \\ 
  \hline
   OA & $\mu$ & $\sigma^2$ & $\mu$ & $\sigma^2$ & $\mu$ & $\sigma^2$ & $\mu$ & $\sigma^2$  \\
  \hline
  Train Data & 0.82 & 0.018 & 0.88 & 0.012 & 0.82 & 0.018 & 0.88 & 0.012 \\
  Test Data & 0.82 & 0.018 & 0.88 & 0.011 & 0.82 & 0.018 & 0.88 & 0.011 \\
  \hline
  Diffusion & \textbf{0.81} & \textbf{0.017} & \textbf{0.85} & \textbf{0.013} & \textbf{0.80} & \textbf{0.021} & \textbf{0.86} & \textbf{0.015} \\
  Autoregression & 0.76 & 0.024 & 0.82 & 0.015 & - & - & - & - \\
  Interpolation & 0.94 & 0.004 & 0.96 & 0.004 & 0.87 & 0.014 & 0.91 & \textbf{0.009} \\
  $\mathcal{N}(0, I)$ Prior   & 0.69 & 0.033 & 0.79 & 0.016 &  0.73 & 0.033 & 0.82 & 0.018\\
  
  \hline
  \end{tabular}
  }
  \end{center}
  \caption{Unnormalized framewise self-similarity (overlapping area) evaluation of unconditional and conditional samples. Evaluations of same samples as in Table~\ref{rel-table}. Note the interpolations have unrealistically high mean overlap and low variance, while the Gaussian prior and TransformerMDN samples suffer from unrealistically lower mean overlap and higher variance.}
  \label{abs-value-table}
\end{table}

\begin{table}[h]
  \begin{center}
  \resizebox{\columnwidth}{!}{%
  \begin{tabular}{|l|c|c|c|c|c|c|c|c|r|}
  \hline
  Setting      & \multicolumn{2}{c}{Unconditional} \vline & \multicolumn{2}{c}{Infilling} \vline \\ 
  \hline
   Metric & FD$\times 10^{-2}$ & MMD$\times 10^{-2}$  & FD$\times 10^{-2}$ & MMD$\times 10^{-2}$  \\
  \hline
  Train Data & 0.00   & 0.00  & 0.00   & 0.00 \\
  Test Data & 1.24   & 0.12  & 1.24   & 0.12 \\
  \hline
  Diffusion & 1.66   & 0.18   & 1.53   & 0.16  \\
  Autoregression & \textbf{1.26}   &\textbf{0.12} & - & - \\
  Interpolation &  3.22  & 0.43   &  1.97  & 0.23   \\  
  $\mathcal{N}(0, I)$ Prior &  2.44  & 0.29  & \textbf{1.17}   & \textbf{0.12} \\  
  \hline
  \end{tabular}
  }
  \end{center}
  \caption{Latent space evaluation of infilling and unconditional and conditional samples. As described in Section~\ref{sec:latent}, the TransformerMDN performs better in latent space similarity, even while producing less realistic samples (as seen in Tables~\ref{rel-table} and \ref{abs-value-table}).}
  \label{latent-table}
\end{table}

\section{Additional Samples}
\begin{figure}[ht]
\vskip 0.2in
\begin{center}
\centerline{\includegraphics[width=\columnwidth]{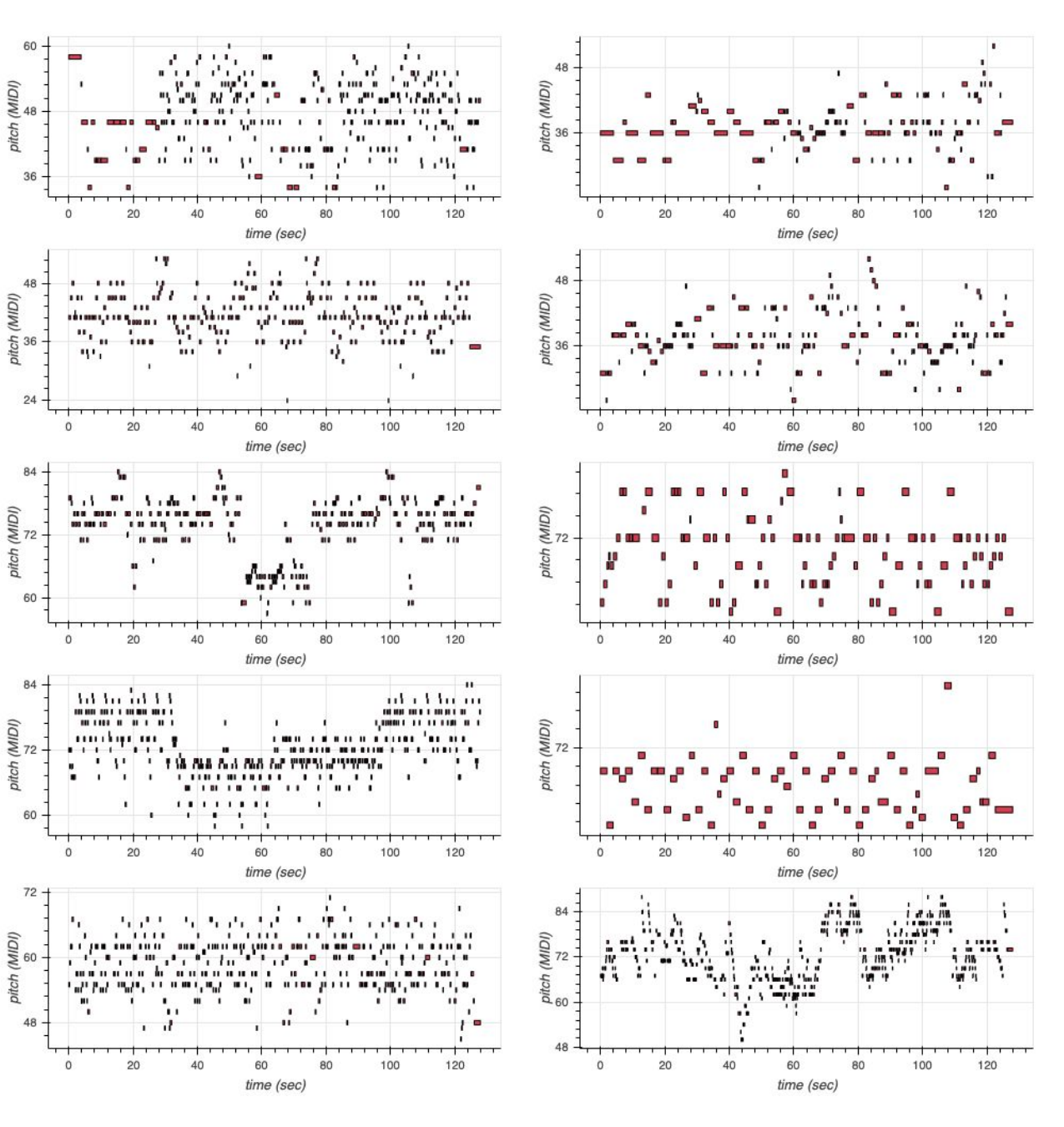}}
\caption{Additional piano rolls from the test set.}
\label{extended-real}
\end{center}
\vskip -0.2in
\end{figure}

\begin{figure}[ht]
\vskip 0.2in
\begin{center}
\centerline{\includegraphics[width=\columnwidth]{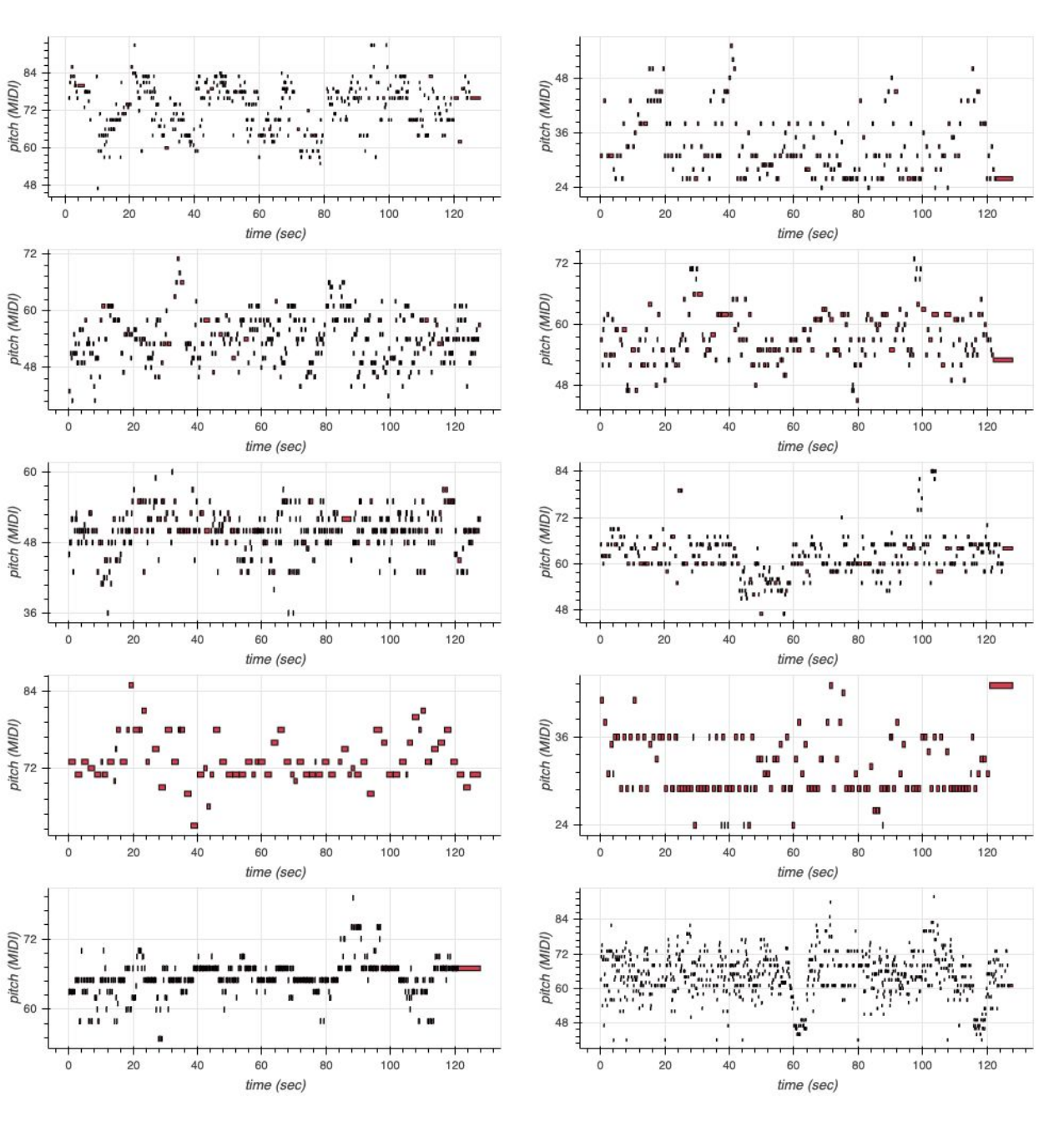}}
\caption{Additional piano rolls generated unconditionally by our diffusion model.}
\label{extended-gen}
\end{center}
\vskip -0.2in
\end{figure}

\begin{figure}[ht]
\vskip 0.2in
\begin{center}
\centerline{\includegraphics[width=\columnwidth]{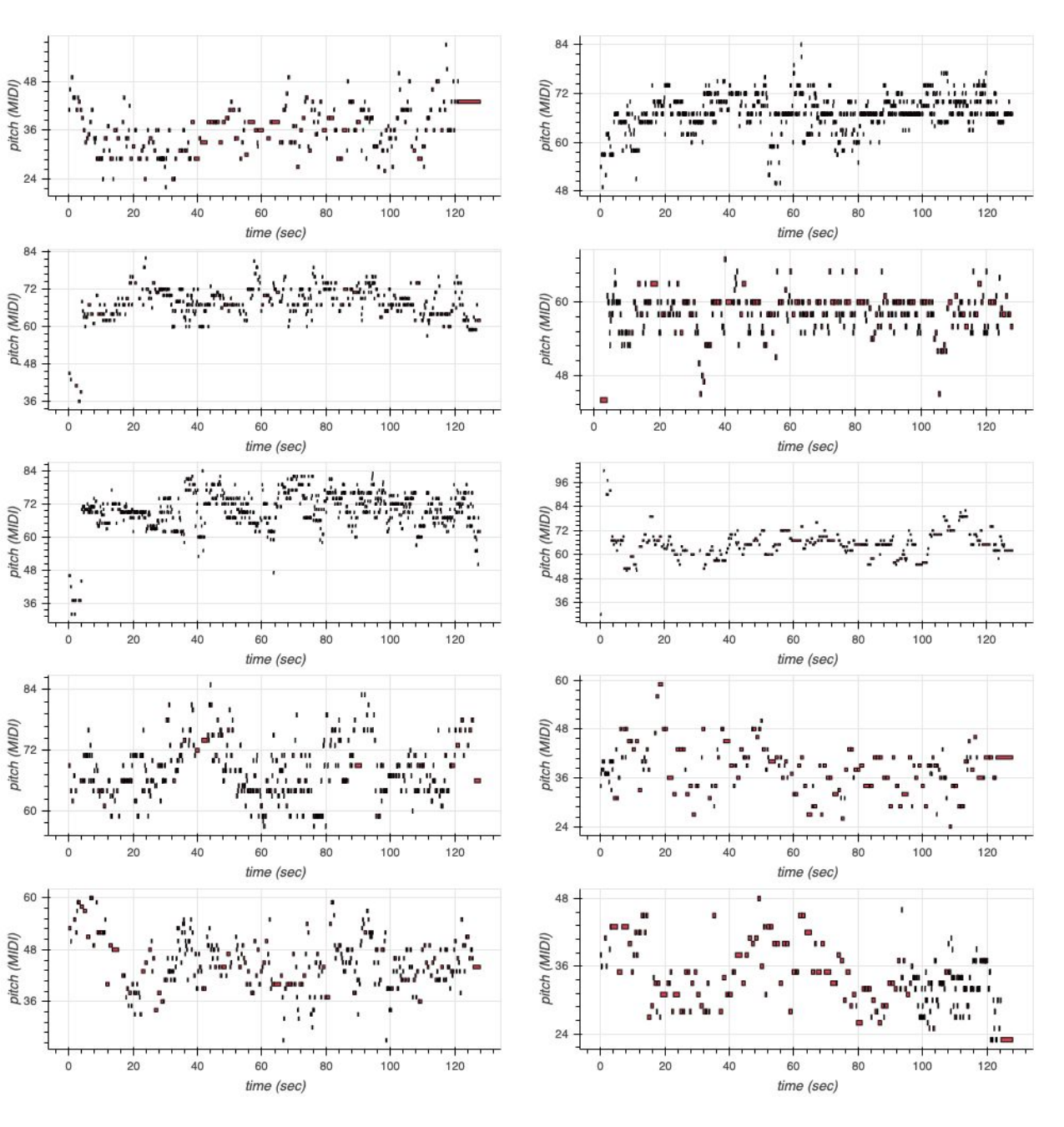}}
\caption{Additional piano rolls generated unconditionally by TransformerMDN.}
\label{extended-mdn}
\end{center}
\vskip -0.2in
\end{figure}

\begin{figure}[ht]
\vskip 0.2in
\begin{center}
\centerline{\includegraphics[width=\columnwidth]{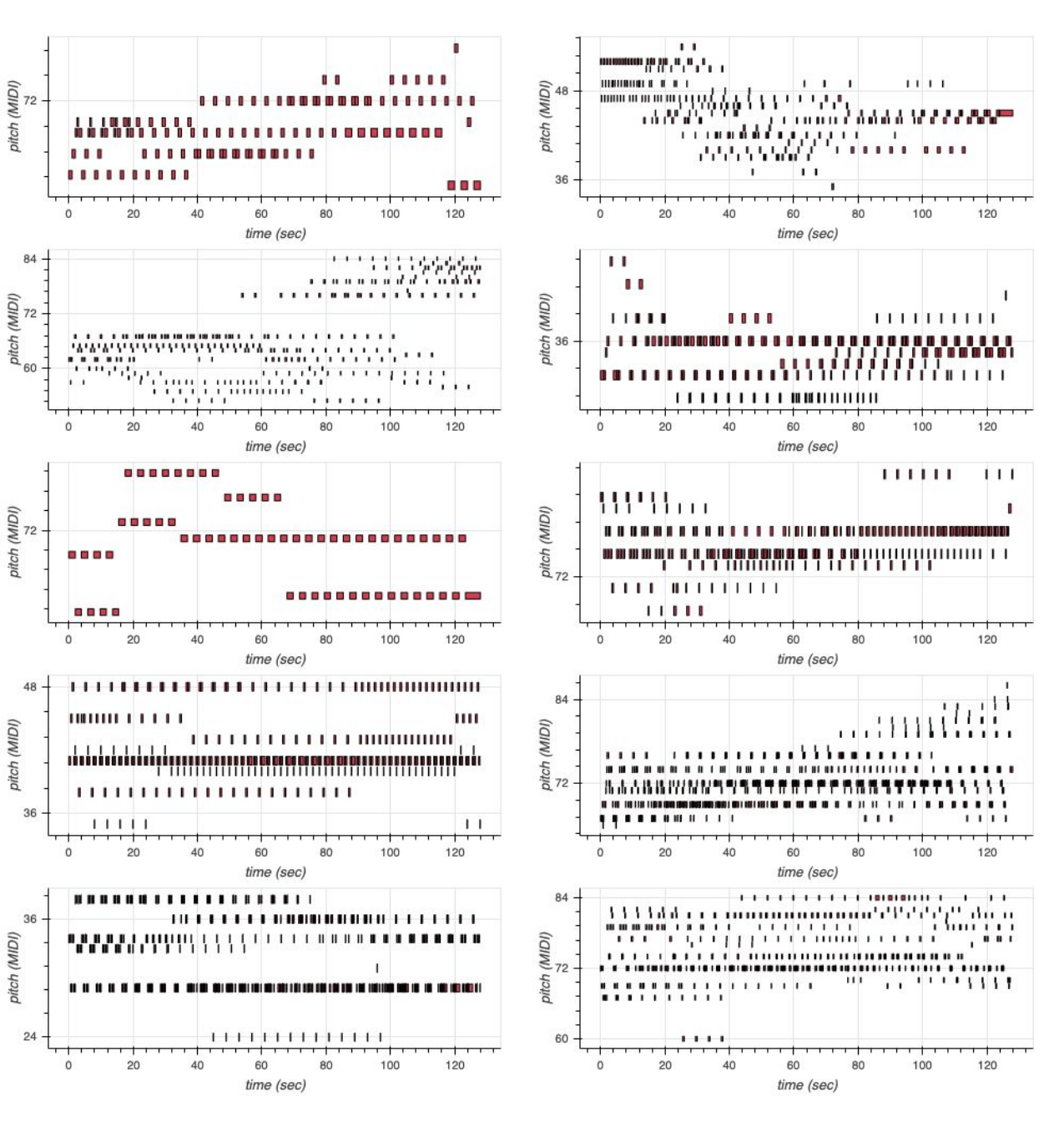}}
\caption{Additional piano rolls generated by performing spherical interpolation~\cite{white2016sampling} between the first and last latent embeddings of sequences drawn from the test set.}
\label{extended-interp}
\end{center}
\vskip -0.2in
\end{figure}

\begin{figure}[ht]
\vskip 0.2in
\begin{center}
\centerline{\includegraphics[width=\columnwidth]{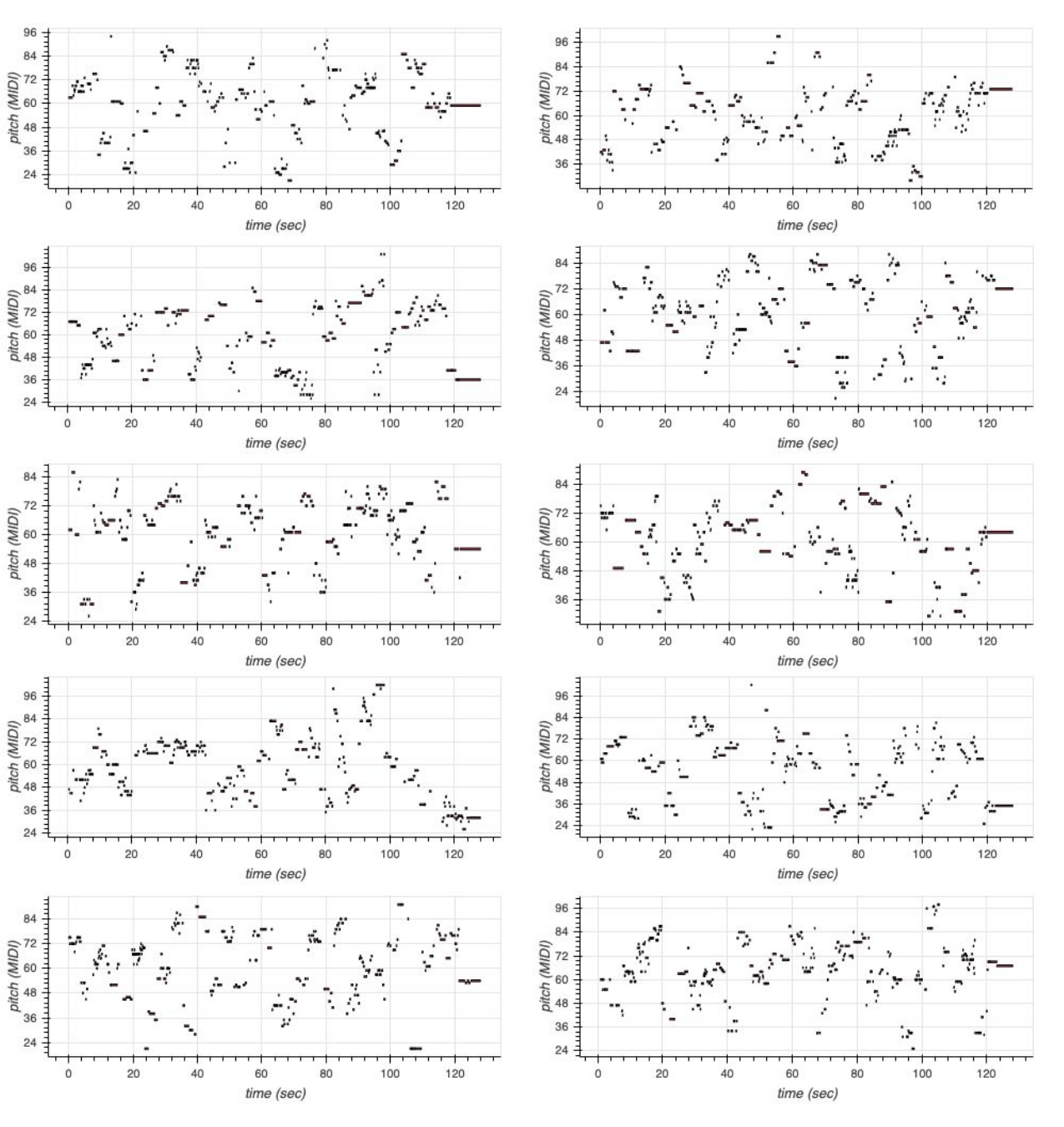}}
\caption{Additional piano rolls generated by sampling each latent embedding independently from the $\mathcal{N}(0, I)$ MusicVAE prior.}
\label{extended-prior}
\end{center}
\vskip -0.2in
\end{figure}

\begin{figure}[ht]
\vskip 0.2in
\begin{center}
\centerline{\includegraphics[width=\columnwidth]{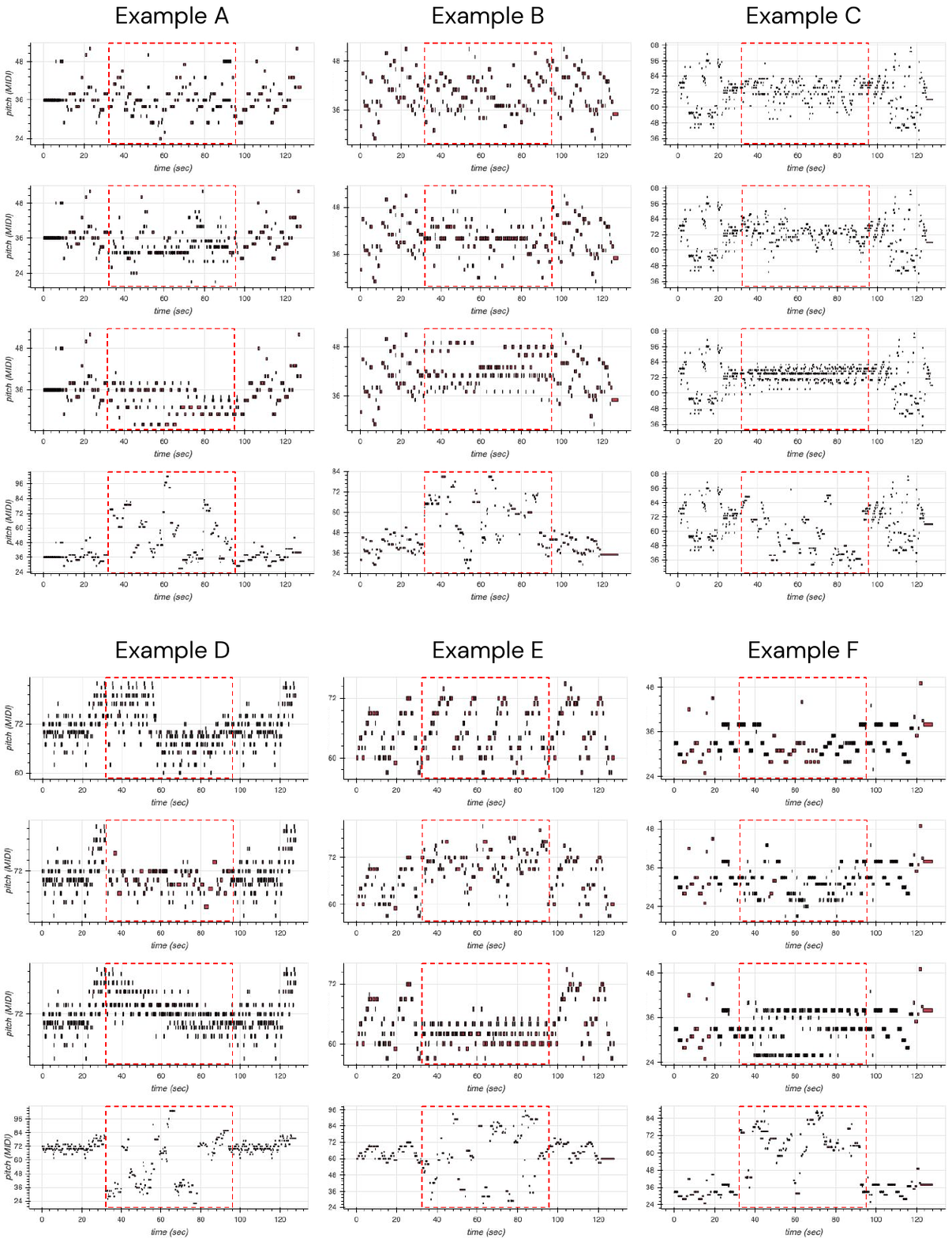}}
\caption{Additional piano rolls of infilling experiments. The first and last 256 melody tokens are held constant and the interior 512 tokens are filled in by the model (dashed red box). Original sample (first row), diffusion model (second row), interpolation (third row), sampling independently from the MusicVAE prior (fourth row).}
\label{extended-infill}
\end{center}
\vskip -0.2in
\end{figure}

In Figure \ref{extended-real} we provide piano rolls of sequences drawn from the test set and in Figures \ref{extended-gen}, \ref{extended-mdn}, \ref{extended-interp}, and \ref{extended-prior} we present additional samples unconditionally generated by our diffusion model, TransformerMDN, spherical interpolation, and through independent sampling from the MusicVAE prior, respectively. Additional piano roll visualizations from infilling experiments are provided in Figure \ref{extended-infill}.

For extended visual and audio samples of the generated sequences from each model, we refer the reader to the online supplement available at \url{https://goo.gl/magenta/symbolic-music-diffusion-examples}.

\onecolumn

\end{document}